\documentclass[prl,twocolumn,showkeys,showpacs,preprintnumbers,amsmath,amssymb]{revtex4}

\usepackage{graphicx}
\usepackage{dcolumn}
\usepackage{bm}
\usepackage{times}

\newcommand{\bea}{\begin{eqnarray}}
\newcommand{\eea}{\end{eqnarray}}
\newcommand{\bean}{\begin{eqnarray*}}
\newcommand{\eean}{\end{eqnarray*}}
\newcommand{\nn}{\nonumber \\}

\def\W #1{\widetilde{#1}}
\def\WH #1{\widehat{#1}}

\def\gb #1{ \left\langle #1 \right]}

\def\a{{\alpha}}

\def\b{{\beta}}

\def\Spab{\gb}

\def\Label#1{\label{#1}%
  \smash{\hbox to0pt{\raise1ex\hbox{\tiny[#1]}\hss}}}

\begin{document}

\preprint{}

\title{ Gauge Amplitude Identities by On-shell Recursion Relation in S-matrix Program}

\author{Bo Feng$~^a$, Rijun Huang$~^b$, Yin Jia$~^b$ }
\affiliation{$~^a$ Center of Mathematical Science, Zhejiang
 University, Hangzhou, China\\ $~^b$Zhejiang Institute of Modern Physics, Physics
Department, Zhejiang University, Hangzhou, China}


\date{\today}

\begin{abstract}

Using only the Britto-Cachazo-Feng-Witten(BCFW) on-shell recursion
relation 
we prove color-order reversed relation, $U(1)$-decoupling relation,
Kleiss-Kuijf(KK) relation
and Bern-Carrasco-Johansson(BCJ) relation
for color-ordered gauge amplitude in the framework of S-matrix
program without relying on Lagrangian description. Our derivation is
the first pure field theory proof of the new discovered BCJ
identity, which substantially reduces the color ordered basis from
$(n-2)!$ to $(n-3)!$. Our proof gives also its physical
interpretation as the mysterious bonus relation with ${1\over z^2}$
behavior under suitable on-shell deformation for no adjacent pair.

\end{abstract}

\maketitle

\section{Introduction}

S-matrix program\cite{S-matrix} is a program to understand the
scattering amplitude of quantum field theory based only on some
general principles, like the Lorentz invariance, Locality,
Causality, Gauge symmetry as well as Analytic property. The
significance of this approach is its generality: results so obtained
do not rely on any detail information of theories, such as   the
Lagrangian description of theories.

However, exactly because its generality with so little assumptions,
there are not much tools available and its study is very
challenging.
One big step along S-matrix program is the unitarity cut method
proposed in \cite{Bern:1994zx}, where on-shell tree amplitudes have
been applied to the calculation of loop amplitudes without  drawing
many many Feynman diagrams.

Another breakthrough is the Britto-Cachazo-Feng-Witten(BCFW)
on-shell recursion relation\cite{Britto:2004ap,Britto:2005fq}. One
way to see it is to
 pick two momenta to do the BCFW-deformation $\WH p_i=p_i+z q$,
$\WH p_j=p_j-zq$ with proper chosen auxiliary null-momentum $q$,
thus the amplitude $A_n$ becomes the analytic function $A_n(z)$ of
$z$ with only single pole structure. By the familiar complex
analysis we can completely determine the function if we know
locations of all poles and their residues. Pole happens when a
propagator reaches mass-shell and the amplitude is effectively
divided into two sub-amplitudes at the left and right (so called
factorization property) of the propagator. Summing up all poles we
obtain
%
{\small \bea A_n=\sum_{{\cal I},{\cal J}} {A_{\cal I}( \WH
p_i(z_{\cal I J}),\WH P^{h}_{\cal I J}(z_{\cal I J} ))A_{\cal J}(
\WH p_j(z_{\cal I J}),-\WH P^{-h}_{\cal I J}(z_{\cal I J} ))\over
P_{\cal I J}^2} ~~~\label{BCFW}\eea}
where sum is over all color-order preserved splitting of
$n$-particles into two on-shell amplitudes with condition $p_i\in
{\cal I}, p_j\in {\cal J}$ while $z_{\cal I J}$ indicates the
particular splitting. The brief form (\ref{BCFW}) is enough for the
understanding of our paper. As reader will see, what we do in whole
paper is to use (\ref{BCFW}) to {\bf expand, recombine and reshuffle
different amplitude components}. It is surprising that with such
simple algebraic manipulations we can get some very deep results.




The derivation of BCFW  relation beautifully demonstrates the idea
of S-matrix program and
its generality has inspired many works, one of them
is the work of Benincasa and Cachazo \cite{Paolo:2007}. In the
paper, by assuming the applicability of BCFW recursion relation they
have  easily re-derived many well known (but difficult to prove)
fundamental facts in S-matrix, such as the Non-Abelian structure for
gauge theory and all matters couple to gravity with same coupling
constant.

In this paper we will focus on  one fundamental object in gauge
theory, i.e., the color-cyclic-ordered tree-level gluon amplitudes,
which are gauge invariant, dynamical building blocks with
Lie-algebra structure having been stripped away. More explicitly we
will use S-matrix framework to discuss following four identities
among these primary amplitudes: (1) Color-order reversed relation $
A(1,2,...,n)=(-)^n A(n,n-1,...,1)$;
(2) The $U(1)$-decoupling relation\cite{Kleiss:1988ne}
$\sum_{\sigma\in cyclic} A_n(1, \sigma(2,3,...,n))=0$;
(3) The  Kleiss-Kuijf relation \cite{Kleiss:1988ne}
%
{\small\bea  A_n(1,\{\a\}, n,\{\b\}) = (-1)^{n_\b}\sum_{\sigma\in
OP(\{\a\},\{\b^T\})} A_n(1,\sigma, n)~.~~~~\label{KK-rel}\eea}
where Order-Preserved(OP) sum is over all permutations of the set
$\a \bigcup \b^T$ where the relative ordering in each set  $\a$  and
$\b^T$ (which is the reversed ordering of set $\b$) is preserved.
The $n_\b$ is the number of $\b$ elements. (4) The
BCJ-relation\cite{Bern:2008qj}
{\small\bea A_n(1,2,\{\a\}, 3, \{ \b\}) & = & \sum_{\sigma_i \in
POP(\{\a,\b\}) } A_n (1,2,3,\sigma_i) {\cal
F}_i~~~~\label{BCJ-general}\eea}
where Partial-Order-Preserved(POP)sum is over all permutations of
set $\a\bigcup \b$ with preserving the relative ordering inside the
set $\b$. The ${\cal F}_i$ are some dynamical factors and explicit
definition can be found in \cite{Bern:2008qj}.

These four identities have been understood from different
perspectives. The properties (1) and (2) can be shown from
Lie-algebra structure. Property (3) is inspired from string theory
and then shown in field theory \cite{DelDuca:1999rs} using different
color decomposition. Property (4) is conjectured through the
Jacobi-identity but has only been proved from string theory
\cite{BjerrumBohr:2009rd} (see further study \cite{Tye:2010dd}).


These four identities, especially the KK and BCJ relations, contain
unexpected important properties of gauge theory. Our proof in
S-matrix frame unifies the treatment of them all and makes them hold
in general ground. Especially our proof is the first pure field
theory proof of BCJ relation. Furthermore our method can be applied
to the field theory understanding of another very important
Kawai-Lewellen-Tye(KLT) relation \cite{Kawai:1985xq,KLT}, which has
only been shown  from string theory. The importance of BCJ and KLT
relations lies in the mysterious observation: on-shell gravity likes
the square of gauge theory while their off-shell Lagrangian
descriptions are completely different (one is normalizable and
another one, unnormalizable). Understanding these observations in
field theory will help us with the searching of consistent quantum
gravity theory, which is still one of most fundamental open problems
in physics.

\section{The Color-order Reversed relation}

One basic observation of \cite{Paolo:2007} is that color-ordered
three particle amplitude is completely fixed by Lorentz symmetry and
satisfy $A(1,2,3)=(-)A(3,2,1)$ without using any Lie-algebra
property. Using the BCFW recursion relation with pair $(n,1)$, we
get
{\small\bean & & A(n,\b_1,.,\b_{n-2},1) \\& = & \sum_{i=1}^{n-3}
A(\WH n,\b_1,.,\b_i,-\WH P_i) {1\over P_i^2} A(\WH
P_i,\b_{i+1},.,\b_{n-2},\WH 1)\nn
& = & \sum_{i=1}^{n-3}(-)^{n-i} A(\WH 1,\b_{n-2},.,\b_{i+1},\WH P_i)
{1\over P_I^2} (-)^{i+2} A(-\WH P_i,\b_i,.,\b_1,\WH n)\nn
& = & (-)^n A(1,\b_{n-2},\b_{n-1},.,\b_1,n)~.\eean}
where we have expanded amplitude  at the second line, then used
induction to reshuffle at the third line and finally recombined at
the fourth line. These manipulations are exactly patterns we will
follow in whole paper. It is worth to notice that we do not need to
specify details like the helicity and shifting of $(n,1)$ as well as
explicit expressions of $A_n$ as long as BCFW on-shell recursion
relation without boundary value is applicable.
Thus our conclusion holds for any helicity configuration.

\section{The $U(1)$-decoupling relation}

The $n=4$ case is easy to check after using the color-reversed
relation in the BCFW expansion. To get more idea of proof, let us
present example of $n=5$ given in (\ref{n=5-U1}). At each line we
use (\ref{BCFW}) to expand left hand side into the right hand side.
To make formula compact we have used, for example $P_{523}$ to
represent amplitude $A(\WH 5,2,3,-\WH P_{523})/s_{523}$, thus $
A(1,4,P_{523})$ really represents $A(\WH 5,2,3,-\WH
P_{523})A(P_{523},1,4)/s_{523}$. By our purposely arrangement, it is
easy to see that the sum of each column at the right hand side is
zero after we use the $U(1)$-decoupling equation for $n=3$ and $n=4$
by induction.

%
\begin{widetext}\vspace{-0.8cm}
{\small\bea \begin{array}{lllllllllll} A(1,2,3,4,5) & = & A(1,
P_{23}, 4,5) &
+ & A(1,P_{234},5) & + & 0 & + & 0 & + & 0 \\
A(1,5,2,3,4) & = & A(1,5,P_{23},4) & + & A(1,5, P_{234}) & + & A(1,
P_{52}, 3,4) & + & A(1, P_{523},4) & + & 0\\
A(1,4,5,2,3) & = & A(1,4,5,P_{23}) & + & 0 & + & A(1,4,P_{52},3) & +
& A(1,4,P_{523}) & + & A(1,P_{452}, 3)\\
A(1,3,4,5,2) & = & 0 & + & 0 & + & A(1,3,4,P_{52}) & + & 0 & + &
A(1,3, P_{452})
\end{array} ~~\label{n=5-U1}\eea}
\end{widetext}\vskip-15pt

Having the experience of $n=5$, the proof for general $n$ by
induction is again by BCFW expanding each amplitude first, then
regrouping every piece into $U(1)$-identity for the lower $m$.
For example, with $(1,2)$-shift the expansion of a
general amplitude
{\small\bea  & & A_n^k(\hat{2},3,\ldots,k,\hat{1},k+1,\ldots,n)\nn
&=&
A_n^{k,0}(\hat{2},3,[4,\ldots,k],\hat{1},k+1,\ldots,n)\nonumber\\
&&+A_n^{k,1}(n,\hat{2},[ 3,\ldots,k],\hat{1},k+1,\ldots,n-1)+\ldots\nonumber\\
&&+A_n^{k,n-k}(k+1,\ldots,n,\hat{2},[ 3,\ldots,k-1],k,\hat{1}),\eea}
where $[i,\ldots,j]$ means sum of all  divisions between legs $i$ to
$j$ and $(k,t)$ means there are $t$ particles in front of $\WH 2$.
It can be checked that with fixed $t$, the sum of $k$ is indeed the
$U(1)$-decoupling identity with lower $m$ and is zero. Having all
possible $t$ we get the identity for $n$, thus finished the proof.

%
%
%
%
%
%
%
%

\section{The KK-relation}

The $n=4$ case is easy to show by BCFW expansion and color-order
reversed relatioin, so again we use the $n=5$ as demonstration of
our proof. With $(1,5)$-shifting we can expand $A(1,2,5,3,4)$ as
{\small\bean &   & A(4,\WH 1,2,\WH P_{35}|-\WH P_{35},\WH 5,3)+
A(3,4,\WH 1,\WH P_{25}|-\WH P_{25}, 2,\WH 5) \nn & &  + A(\WH
1,2,-\WH P_{12}|\WH P_{12},\WH 5,3,4) + A(4,\WH 1,-\WH P_{41}|\WH
P_{41}, 2,\WH 5,3) \eean}
Then we use the KK-relation for the four gluon part, i.e., $A(4,\WH
1,2,\WH P_{35})=-A(\WH 1,2,4,\WH P_{35})-A(\WH 1,4,2,\WH P_{35})$ as
well as the one for $A(\WH P_{41}, 2,\WH 5,3)$ to get (notice we
have used the color-order reverse relation)
{\small\bean  &  & A( (1,2,4,P_{35})+(1,4,2,P_{35})|3,5)  +
A(1,4,3,P_{25}| 2,5)\nn & & + A(1,2|P_{12},
4,3,5)+A(1,4|(P_{41},2,3,5)+(P_{41},3,2,5))\eean}
It is easy to see that among these six terms,  $T_1+
T_4=A(1,2,4,3,5)$, $T_2+T_5=A(1,4,2,3,5)$ and
$T_3+T_6=A(1,4,3,2,5)$, thus by the recombination
 we have produced
the KK-relation for $A(1,2,5,3,4)$.

Having above example,
the proof of the general case $A(1,\{\a_1,.,\a_k\},
n,\{\b_1,.,\b_m\})$ with $(1,n)$-shifting is done first by expanding
as
\begin{widetext}\vspace{-0.8cm}
{\small\bea & & A(1,\{\a_1,,\a_k\}, n,\{\b_1,.,\b_m\}) =
\left[\sum_{i=0}^k \sum_{j=0}^m A(\b_{j+1},.,\b_m,1,\a_1,.,\a_i,
P_{ij}|-P_{ij}, \a_{i+1},.,\a_k, n,\b_1,.,\b_j) \right]_{(i,j)\neq
(0,m),(k,0)}~~~~\label{KK-gen-left}\eea}
\end{widetext}\vskip-15pt
where two cases $(i=0,j=m)$ and $(i=k,j=0)$ should be excluded from
the summation. Now we use the induction for each component, i.e.,
$A(\b_{j+1},.,\b_m,1,\a_1,.,\a_i, P_{ij})  = (-)^{m-j}
\sum_{\sigma_{ij}} A(1, \sigma_{ij}, P_{ij})$ and similarly for the
second factor.
With some calculations like previous example of $n=5$, it is easy to
see that for each given set $\{i,j,\sigma_{ij},\W\sigma_{ij}\}$,
(\ref{KK-gen-left}) gives a term at the right hand side of
(\ref{KK-rel})  with legitimate ordering and BCFW splitting. Thus if
we can show that number of terms for both expansions are same, the
proof is done.


To count terms, it is easy to see that there are $C^i_{i+m-j}$ and
$C^j_{j+k-i}$ terms for each factor  respectively at the right hand
side of (\ref{KK-gen-left}). Thus the total number of terms at the
right hand side of (\ref{KK-gen-left}) is
{\small\bea -2 {(m+k)!\over m! k!}+\sum_{i=0}^k \sum_{j=0}^m
{(i+m-j)!\over i! (m-j)!}{(j+k-i)!\over j! (k-i)!}\eea}
where $-2{(m+k)!\over m! k!}$ counts the two excluded cases. The
right hand side of KK-relation (\ref{KK-rel}) will be ${(k+m)!\over
k! m!} (k+m-1)$ after we have used the BCFW to expand each amplitude
into $(k+m-1)$ terms. These two numbers match up as it should be.

\section{The BCJ relation}

The BCJ relation (\ref{BCJ-general}) is more complicated since the
appearance of various dynamical factors $s_{ij}$. In its most
general form, the set $\a,\b$ can be arbitrary. However, we want to
show that all other equations are redundant except the one where the
set $\a$ has only one element, which we call  the "fundamental
BCJ-relation". More accurately we want to show that if these
fundamental BCJ-relations are true, combining with $U(1)$-decoupling
relation and KK-relation we can express any amplitude by $(n-3)!$
amplitudes of the form $A(1,2,3,\sigma(4,.n))$. This is exact the
same statement given by general BCJ-relation.

To show that, let us start from the configuration $A(1,2,\{ t_1\},
t_2,\{ t_3,..., t_{n-3}, 3\}$, i.e., the particle $3$ is at the
location $n$ at the left hand side of the fundamental BCJ-relation.
By the expansion at the right hand side of BCJ-relation, particle
$3$ will have two locations at each equation: one is at the location
$n$ and one is at the location $(n-1)$. There are $(n-3)!$
equations, thus we can use them to solve all configurations of $3$
at the location $n$ by the one at the location $(n-1)$. At the next
step we consider the configuration at the left hand side of
fundamental BCJ-relation with $3$ at the location $(n-1)$. By the
expansion of the BCJ-relation at the right hand side we see now that
$3$ can be located at $(n-1)$ and $(n-2)$, thus we can solve $3$ at
the location $(n-1)$ by the one at location $(n-2)$. Iterating this
procedure we can solve $3$ at the location $5$ by the one at the
location $4$ and finally we solve the one at the location $4$  by
the one at the location $3$.

Now let us write down the form of fundamental BCJ-relation for
$n=4,5,6$ as following:
{\small\bea 0 & = & I_4=A(2,4,3,1)(s_{43}+s_{41})+A(2,3,4,1)
s_{41}\nn
0 & = & I_5=A(2,4,3,5,1)(s_{43}+s_{45}+s_{41})\nn & &
+A(2,3,4,5,1)(s_{45}+ s_{41})+A(2,3,5,4,1)s_{41}\nn
0 & = & I_6=A(2,4,3,5,6,1)(s_{43}+s_{45}+s_{46}+s_{41}) \nn & &
+A(2,3,4,5,6,1)(s_{45}+s_{46}+s_{41})\nn & &
+A(2,3,5,4,6,1)(s_{46}+s_{41})+ A(2,3,5,6,4,1)s_{41}\eea}
and obviously generalization for general $n$. There are  two
observations useful later. The first one is the special relation for
$n=3$, i.e, $A(2,3,1) s_{31}=0$. The second one is that we can use
momentum conservation to write above relation into dual form, for
example, the case $n=5$ can be rewritten as
{\small\bean 0 & = & A(2,4,3,5,1) s_{24}+A(2,3,4,5,1)(s_{24}+
s_{34})\nn & & +A(2,3,5,4,1)(s_{24}+s_{34}+s_{54})\eean}

Before we present our general proof by induction, let us consider
how we can derive the BCJ-relation for $n=4$. Starting from the
$U(1)$-decoupling with $(1,2)$-shifting
we consider following contour
integration expression which is zero by $U(1)$-decoupling relation
{\small\bea \oint {dz\over z} s_{\WH 23}(z) [ A(\WH 1, \WH 2,
3,4)+A(\WH 1, 3,4,\WH 2)+A(\WH 1, 4,\WH 2,
3)]=0~.~~~\label{BCJ-n=4-contour}\eea}
Among these three terms,  since the multiplication of factor $s_{\WH
23}(z)$, $A(\WH 1, \WH 2,3,4)$ has only one pole contribution at
$z=0$, thus we have $T_1= s_{23} A(1,2,3,4)$. The third term is
zero, since $\WH 1, \WH 2$ are not nearby and the large $z$ limit of
amplitude is in fact ${1\over z^2}$. The second term is given by
$T_2=(s_{23}-s_{23}(z_{13})) A(1,3,4,2)=-s_{13} A(1,3,4,2)$.
%
%
Putting all results together and using the color-reserved relation
we get immediately $s_{23} A(2,3,4,1)+(s_{23}+s_{43}) A(2,4,3,1)=0$.
%
%

Having done for $n=4$, we move to the general proof using the
induction. To make the step clear, we consider the case $n=6$ and
arbitrary $n$ is easily dealt with same method. Taking the
$\Spab{2|1}$-shifting and using the BCFW recursion relation to
expand each amplitude in $I_6$,
 we will get three different splitting for each amplitude. Let us
consider the splitting
{\small\bean I_6^{[2]}&= & A(\WH 2,4,-\WH P_{24}|\WH
P_{24},3,5,6,\WH 1)(s_{43}+s_{45}+s_{46}+s_{41}) \nn & & + A(\WH
2,3,-\WH P_{23}|\WH P_{23},4,5,6,\WH 1)(s_{45}+s_{46}+s_{41})\nn
& & +A(\WH 2,3,-\WH P_{23}|\WH P_{23},5,4,6,\WH 1)
(s_{46}+s_{41})\nn & & +A(\WH 2,3,-\WH P_{23}|\WH P_{23},5,6,4,\WH
1)s_{41}\eean}
where the splitting parameter $[2]$ means  there are two particles
at the left hand side.  All terms of $I_6^{[2]}$ can be divided into
two categories: the one with $4$ at the left hand side and the
another one, right hand side. The last three terms with $4$ at the
right hand side can be rewritten as
{\small\bean & &  A(\WH 2,3,-\WH P_{23}|\WH P_{23},4,5,6,\WH
1)(s_{45}+s_{46}+s_{4\WH 1}) \nn & & +A(\WH 2,3,-\WH P_{23}|\WH
P_{23},5,4,6,\WH 1) (s_{46}+s_{4\WH 1})\nn & & +A(\WH 2,3,-\WH
P_{23}|\WH P_{23},5,6,4,\WH 1)s_{4\WH 1}\nn
& & + \left\{A(\WH 2,3,-\WH P_{23}|\WH P_{23},4,5,6,\WH 1) +A(\WH
2,3,-\WH P_{23}|\WH P_{23},5,4,6,\WH 1) \right.\nn & & \left.+A(\WH
2,3,-\WH P_{23}|\WH P_{23},5,6,4,\WH 1)\right\}( s_{41}-s_{4\WH
1}(z_{23}))\eean}
By the induction over the second factor we know the sum of first
three lines are zero. The first term of $I_6^{[2]}$ can be rewritten
in dual form as
{\small\bean & &  -s_{24}A(\WH 2,4,-\WH P_{24}|\WH P_{24},3,5,6,\WH
1)\nn & = & -s_{\WH 24}(z_{24})A(\WH 2,4,-\WH P_{24}|\WH
P_{24},3,5,6,\WH 1) \nn
& &  -(s_{24}-s_{\WH 2 4}(z_{24}))A(\WH 2,4,-\WH P_{24}|\WH
P_{24},3,5,6,\WH 1)\eean}
where again the first term is zero by induction. Using the fact
$-(s_{24}-s_{\WH 2 4}(z_{24}))= ( s_{41}-s_{4\WH 1}(z_{24}))$, we
can put them together as
{\small \bean & & I_6^{[2]}=  ( s_{41}-s_{4\WH 1}(z_{23}))\left\{
A(\WH 2,3,-\WH P_{23}|\WH P_{23},4,5,6,\WH 1) \right. \nn
& & \left. +A(\WH 2,3,-\WH P_{23}|\WH P_{23},5,4,6,\WH 1) +A(\WH
2,3,-\WH P_{23}|\WH P_{23},5,6,4,\WH 1)\right\}\nn & & +A(\WH
2,4,-\WH P_{24}|\WH P_{24},3,5,6,\WH 1)( s_{41}-s_{4\WH
1}(z_{24}))\eean}
By similar manipulations for $I_6^{[3]}, I_6^{[4]}$ and sum all
three together we finally have
{\small\bea I_6&= & s_{41}\left\{ A(2,4,3,5,6,1) +
A(2,3,4,5,6,1)\right. \nn & & \left.+A(2,3,5,4,6,1)
+A(2,3,5,6,4,1)\right\}\nn & & +\oint_{z\neq 0} {dz s_{\WH 1 4}\over
z}\left\{ A(\WH 2,4,3,5,6,\WH 1) + A(\WH 2,3,4,5,6,\WH 1)\right. \nn
& & \left.+A(\WH 2,3,5,4,6,\WH 1)+A(\WH 2,3,5,6,4,\WH
1)\right\}\eea}
where the contour integration has excluded the contribution at the
pole $z=0$.  Using the KK-relation, we can rewrite it as
{\small\bea -I_6&= & s_{41} A(4,2,3,5,6,1) +\oint_{z\neq 0} {dz
s_{\WH 1 4}\over z} A(4,\WH 2,3,5,6,\WH 1)~~~~\eea}
Now noticing that $(1,2)$ are not nearby, thus $\oint {dz s_{\WH 1
4}\over z} A(4,\WH 2,3,5,6,\WH 1)=0$ by the ${1\over z^2}$ behavior
at infinity, and we get
{\small\bean & & \oint_{z\neq 0} {dz s_{\WH 1 4}\over z}
A(4,2,3,5,6,1)=-\oint_{z= 0} {dz s_{\WH 1 4}\over z} A(4,2,3,5,6,1)
\nn & = & -s_{41} A(4,2,3,5,6,1)\eean}
Putting it back  we have finally proved $I_6=0$.

The proof for general $n$ will be exact same as the one with $I_6$
and given by $s_{41} A_n(4,2,3,5,..,n,1)-\oint_{z=0}{dz s_{\WH 1
4}\over z}A_n(4,\WH 2,3,5,..,n,\WH 1)=0$.


Let us have some final remarks. In the proof of BCJ relation, it is
crucial that when shifted pair $(i,j)$ are not nearby, there is a
deformation making the amplitude vanishing as ${1\over z^2}$. With
this better behavior we should have some bonus  relation as found in
gravity in \cite{ArkaniHamed:2008gz,Spradlin:2008bu}. From this
paper, now we know the mysterious bonus relation in gauge theory is
nothing, but the BCJ-relation.

The BCJ-relation has not been explored extensively, but its
potential importance is manifest. It can be used to speed up
amplitude calculation. Furthermore, its generalization to higher
loop \cite{Bern:2010ue} and its relation to gravity
\cite{Vanhove:2010nf} make it important for the discussion of
finiteness of ${\cal N}=8$ supergravity.


{\bf Acknowledgments:} { We will like to thank N.E.J Bjerrum-Bohr,
P.H. Damgaard and T. Sondergaard for valuable discussion and the
hospitality of Niels Bohr Institute where the final stage of the
work is done. We would also like to thank R. Britto, Y.H He for many
suggestions of the improvement of presentation. The work is funded
by Qiu-Shi funding as well as the Fundamental Research Funds for the
Central Universities with contract number 2009QNA3015, and Chinese
NSF funding under contract No.10875104.}


\begin{thebibliography}{99}







\bibitem{S-matrix} D.I. Olive, Phys. Rev. 135,B 745(1964); G.F.
Chew, "The Analytic S-Matrix: A Basis for Nuclear Democracy",
W.A.Benjamin, Inc., 1966; R.J. Eden, P.V. Landshoff, D.I. Olive,
J.C. Polkinghorne, "The Analytic S-Matrix", Cambridge University
Press, 1966.


\bibitem{Bern:1994zx}
  Z.~Bern, L.~J.~Dixon, D.~C.~Dunbar and D.~A.~Kosower,
  Nucl.\ Phys.\  B {\bf 425}, 217 (1994)
  [arXiv:hep-ph/9403226].


\bibitem{Britto:2004ap}
  R.~Britto, F.~Cachazo and B.~Feng,
  Nucl.\ Phys.\  B {\bf 715}, 499 (2005)
  [arXiv:hep-th/0412308].

\bibitem{Britto:2005fq}
  R.~Britto, F.~Cachazo, B.~Feng and E.~Witten,
  Phys.\ Rev.\ Lett.\  {\bf 94}, 181602 (2005)
  [arXiv:hep-th/0501052].

\bibitem{Paolo:2007}
  Paolo.~Benincasa,   Freddy.~Cachazo,
  arXiv:0705.4305[hep-th].

\bibitem{Kleiss:1988ne}
  R.~Kleiss and H.~Kuijf,
  Nucl.\ Phys.\  B {\bf 312}, 616 (1989).

\bibitem{Bern:2008qj}
  Z.~Bern, J.~J.~M.~Carrasco and H.~Johansson,
  Phys.\ Rev.\  D {\bf 78}, 085011 (2008)
  [arXiv:0805.3993 [hep-ph]].


\bibitem{DelDuca:1999rs}
  V.~Del Duca, L.~J.~Dixon and F.~Maltoni,
  Nucl.\ Phys.\  B {\bf 571}, 51 (2000)
  [arXiv:hep-ph/9910563].


\bibitem{BjerrumBohr:2009rd}
  N.~E.~J.~Bjerrum-Bohr, P.~H.~Damgaard and P.~Vanhove,
  Phys.\ Rev.\ Lett.\  {\bf 103}, 161602 (2009)
  [arXiv:0907.1425 [hep-th]].


  S.~Stieberger,
  arXiv:0907.2211 [hep-th].




\bibitem{Tye:2010dd}
  H.~Tye and Y.~Zhang,
  arXiv:1003.1732 [hep-th].
  N.~E.~J.~Bjerrum-Bohr, P.~H.~Damgaard, T.~Sondergaard and P.~Vanhove,
  arXiv:1003.2403 [hep-th].
T.~Sondergaard,
0903.5453 [hep-th].
  C.~R.~Mafra,
  JHEP {\bf 1001}, 007 (2010)
  [arXiv:0909.5206 [hep-th]].

\bibitem{Kawai:1985xq}
  H.~Kawai, D.~C.~Lewellen and S.~H.~H.~Tye,
  Nucl.\ Phys.\  B {\bf 269}, 1 (1986).

%
\bibitem{KLT}
N.~E.~J.~Bjerrum-Bohr, P.~H.~Damgaard, B.~Feng, T.~Sondergaard,
arXiv:1005.4367, arXiv:1006.3214,  arXiv:1007.3111

\bibitem{ArkaniHamed:2008gz}
  N.~Arkani-Hamed, F.~Cachazo and J.~Kaplan,
  arXiv:0808.1446 [hep-th].


\bibitem{Spradlin:2008bu}
  M.~Spradlin, A.~Volovich and C.~Wen,
  Phys.\ Lett.\  B {\bf 674}, 69 (2009)
  [arXiv:0812.4767 [hep-th]].

\bibitem{Bern:2010ue}
  Z.~Bern, J.~J.~M.~Carrasco and H.~Johansson,
  arXiv:1004.0476 [hep-th].


\bibitem{Vanhove:2010nf}
  P.~Vanhove,
  arXiv:1004.1392 [hep-th].
  Z.~Bern, T.~Dennen, Y.~t.~Huang and M.~Kiermaier,
  arXiv:1004.0693 [hep-th].

\end{thebibliography}
\end{document}